\documentclass[prl,aps,twocolumn,showpacs,
]{revtex4}
\usepackage{amsmath,amssymb,graphicx}
\usepackage{pxfonts}
\usepackage{graphicx}
\usepackage{color}

\begin{document}

\date{\today}

\title{Mott transition in excitonic Bose polarons}

\author{E.~A.~Szwed\textsuperscript{a}}
\affiliation{\textsuperscript{a}Department of Physics, University of California San Diego, La Jolla, CA 92093, USA}
\author{B.~Vermilyea\textsuperscript{a}}
\affiliation{\textsuperscript{a}Department of Physics, University of California San Diego, La Jolla, CA 92093, USA}
\author{D.~J.~Choksy\textsuperscript{a}}
\affiliation{\textsuperscript{a}Department of Physics, University of California San Diego, La Jolla, CA 92093, USA}
\author{Zhiwen~Zhou\textsuperscript{a}}
\affiliation{\textsuperscript{a}Department of Physics, University of California San Diego, La Jolla, CA 92093, USA}
\author{M.~M.~Fogler\textsuperscript{a}}
\affiliation{\textsuperscript{a}Department of Physics, University of California San Diego, La Jolla, CA 92093, USA}
\author{L.~V.~Butov\textsuperscript{a}
} 
\affiliation{\textsuperscript{a}Department of Physics, University of California San Diego, La Jolla, CA 92093, USA}
\author{K.~W.~Baldwin\textsuperscript{b}}
\affiliation{\textsuperscript{b}Department of Electrical Engineering, Princeton University, Princeton, NJ 08544, USA}
\author{L.~N.~Pfeiffer\textsuperscript{b}}
\affiliation{\textsuperscript{b}Department of Electrical Engineering, Princeton University, Princeton, NJ 08544, USA}

\begin{abstract}
\noindent
For a neutral system of positive and negative charges, such as atoms in a crystal, increasing the density causes the Mott transition from bound electrons to free electrons. The density of optically generated electron-hole systems can be controlled in situ by the power of optical excitation that enables the Mott transition from excitons, the bound pairs of electrons and holes, to free electrons and holes with increasing density. These Mott transitions occur in systems of pairs of the same kind, such as atoms or excitons. However, a different type of the Mott transition can occur for Bose polarons. A Bose polaron is a mobile particle of one kind surrounded by a Bose gas of particles of another kind. For the Mott transition in polarons, the polaron states vanish with increasing density of the surrounding gas. In this paper, we present the observation of this type of the Mott transition and the measurement of the Mott transition parameter $n_{\rm M}^{1/2} a_{\rm B}$ in 2D excitonic Bose polarons.  
\end{abstract}
\maketitle

The Mott transition is an interaction-induced transition from a non-metallic state of neutral bound pairs of charged particles to a metallic state of free charged particles~\cite{Mott1961}. In three dimensional systems increasing the density causes the transition to a metallic state when the Mott transition parameter $n_{\rm M}^{1/3}a_{\rm B} \sim 0.25$, where $n_{\rm M}$ is the density of the Mott transition, $a_{\rm B} = \hbar^2 \kappa / m e^2$ the Bohr radius of the pair, $\kappa$ the background dielectric constant, and $m$ the reduced effective mass of the particles in the pair~\cite{Mott1961}. For an array of hydrogen atoms, free electrons form at high densities when the distance between the atoms become comparable to the Bohr radius. The Mott transition at high densities, comparable to the densities of atoms in crystal lattices, is explored in materials~\cite{Mott1968}.

Semiconductors offer an opportunity to explore the Mott transition at lower densities. For donors and acceptors in semiconductors, $\kappa$ should be the dielectric constant of the semiconductor and $m$ should be an effective mass of an electron $m_{\rm e}$ in the case of donors or hole $m_{\rm h}$ in the case of acceptors. The large $\kappa$ and small electron and hole effective masses in a semiconductor allow achieving the Mott transition when the distance between donors (or acceptors) is comparable to the Bohr radius of the bound state of the electron on the donor (or the hole on the acceptor), that is at the density significantly lower than the density of atoms in the crystal lattice. Changing the doping concentration allows implementing the Mott transition in semiconductors~\cite{Mott1961} .

The density of optically generated electron-hole (e-h) systems in semiconductors $n \sim P_{\rm ex} \tau$ is controlled by the power of optical excitation $P_{\rm ex}$ within orders of magnitude ($\tau$ is the e-h recombination lifetime). This in-situ density control enables exploring the Mott transition from excitons, the bound e-h pairs, to e-h plasma of free electrons and holes. The photoluminescence (PL) lineshape allows distinguishing the excitons from the e-h plasma. For instance, for a cold two-dimensional (2D) e-h plasma, the PL spectrum is step-like (due to the step-like 2D density of states) and the PL linewidth is close to the sum of the electron and hole Fermi energies $E_n = \pi \hbar^2 n / \mu_{\rm e-h}$, where $\mu_{\rm e-h} = (m_{\rm e}^{-1} + m_{\rm h}^{-1})^{-1}$ is the reduced e-h mass. For the e-h plasma cooled to the temperature of BCS-like exciton condensation the Cooper-pair like excitons emerge~\cite{Keldysh1965, Comte1982} that enhances the PL intensity at the Fermi edge, keeping however the overall PL linewidth close to $E_n$~\cite{Choksy2023}. In contrast, the PL linewidth of hydrogen-like excitons at low densities in low-disorder materials is significantly smaller than the PL linewidth of e-h plasma~\cite{Keldysh1968}. Therefore, the Mott transition from excitons to e-h plasma with increasing density is revealed by a strong enhancement of the PL linewidth~\cite{Butov1991, Kappei2005, Wang2019}. The Mott transition in optically generated e-h systems is achieved when the distance between excitons is comparable to the exciton Bohr radius. Similar to the case of donors and acceptors, the densities of the Mott transition in e-h systems are significantly lower than the densities of atoms in the crystal lattices. These e-h densities are achieved with optical generation~\cite{Butov1991, Kappei2005, Wang2019}.

The Mott transition in arrays of atoms, doped semiconductors, and optically generated electrons and holes is induced by interaction in a system of pairs of the same kind. Bose polarons give an opportunity to realize a different type of the Mott transition. Bose polarons are mobile particles of one kind dressed by excitations of a degenerate Bose gas of particles of another kind. Bose polarons are explored in Bose gases of ultracold atoms~\cite{Hu2016, Jorgensen2016, Camargo2018, Yan2020, Skou2021, Grusdt2024}, polaritons~\cite{Levinsen2019, Tan2023, Grusdt2024}, and excitons~\cite{Amelio2023, Szwed2024}. For the Mott transition in polarons, the polaron states formed by the particles of one kind vanish with increasing density of the particles of another kind. Achieving this Mott transition with ultracold atoms is challenging due to the required high densities, comparable to the densities of atoms in crystal lattices, that is beyond the range of densities for dilute gases of ultracold atoms~\cite{Cornell2002, Ketterle2002}. However, this type of the Mott transition can be achieved with excitons. In this work, we present the observation of this type of the Mott transition with excitonic Bose polarons. 

\begin{figure}
\begin{center}
\includegraphics[width=8.5cm]{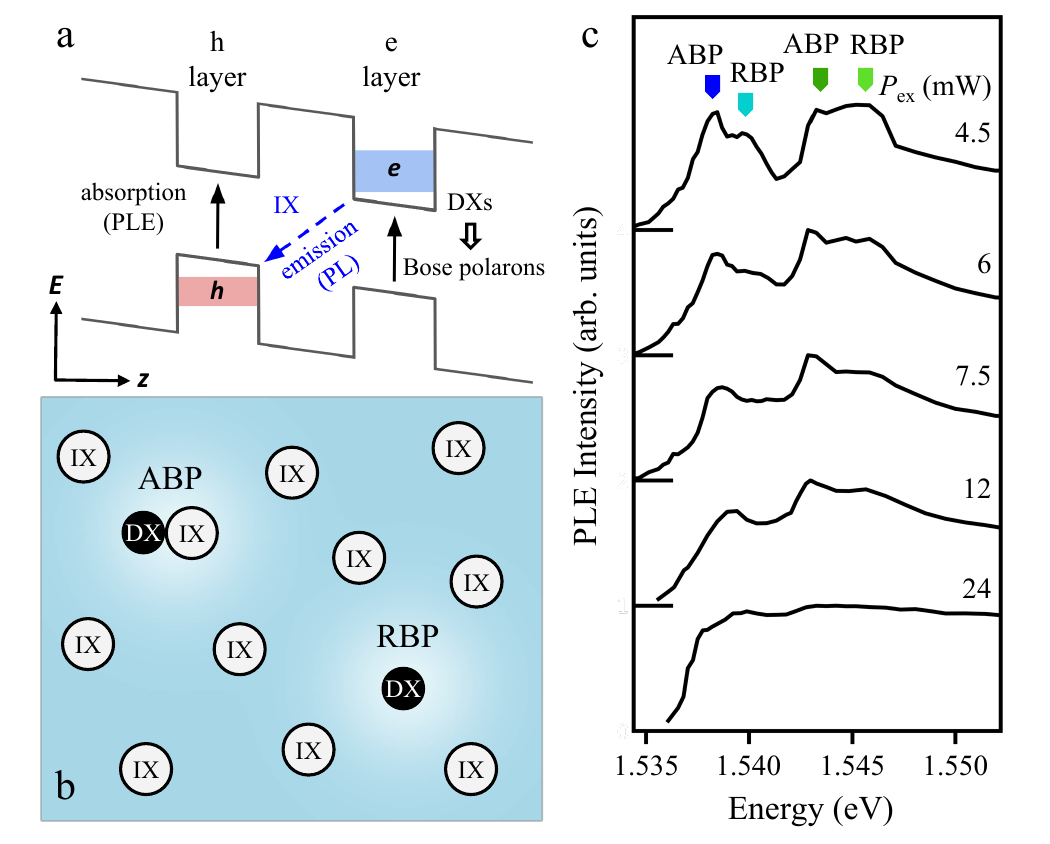}
\caption{Bose polaron diagrams and PLE spectra. 
(a) Diagram of e-h bilayer. Electrons (e) and holes (h) are confined in separate layers. Spatially direct absorption (PLE) is indicated by black arrows, spatially indirect emission (PL) by a blue dashed arrow. DXs and IXs are direct and indirect excitons. 
(b) DX-IX biexcitons and DXs interact with surrounding IXs and form attractive and repulsive Bose polarons (ABP and RBP).
(c) PLE spectra vs. excitation power $P_{\rm ex}$. $T = 10$~K. The two lower-energy peaks (marked by blue arrows) correspond to ABP and RBP for DXs containing heavy holes. The two higher-energy peaks (green arrows) correspond to ABP and RBP for DXs containing light holes. The polaron peaks vanish at high density.
}
\end{center}
\end{figure}

We realize excitonic Bose polarons in e-h bilayers (Fig.~1a). The electron and hole layers are formed by two 15~nm-thick GaAs quantum wells (QWs) separated by a 4~nm-wide AlGaAs barrier in a GaAs/AlGaAs heterostructure. Electrons and holes are optically generated and are driven into the electron and hole layers by an applied voltage. Due to the separation between the electron and hole layers, the lifetime of spatially indirect excitons (IXs), also known as interlayer excitons, exceeds the lifetime of spatially direct excitons (DXs), also known as intralayer excitons, by orders of magnitude, reaching $\tau \sim 1$~$\mu$s in the heterostructure. The long lifetimes allow the optically generated e-h system to cool down to the temperature of the crystal lattice~\cite{Choksy2023}. The density 
of separated electrons and holes $n \sim P_{\rm ex} \tau$ is controlled by the optical excitation power $P_{\rm ex}$. The long lifetimes allow using low $P_{\rm ex}$ that don't overheat the system. The heterostructure details and optical measurements are described in supporting information (SI).

The absorption in e-h bilayers is dominated by spatially direct (intralayer) optical transitions such as the transitions generating DXs~\cite{Szwed2024}. Due to the much lower overlap of the wave functions of separated electrons and holes, absorption via spatially indirect (interlayer) transitions is negligibly weak. The energy relaxation of the optically generated electrons and holes forms the spatially separated electron and hole layers. Due to a fast interlayer relaxation of electrons and holes and long lifetime of IXs, the IX density is much higher than the DX density~\cite{Choksy2021}.

DXs immersed in Bose gases of IXs form Bose polarons. Both attractive and repulsive Bose polarons, ABP and RBP, were observed in recent experiments~\cite{Szwed2024}. At low IX densities, the ABP evolves into a DX-IX biexciton and RBP evolves into a free DX. The energy splitting between ABP and RBP was analyzed in Ref.~\cite{Szwed2024} and was found to be close to the DX-IX biexciton binding energy in the low density limit and enhancing with increasing IX density, in agreement with theoretical calculations~\cite{Szwed2024}.

In this work, we detect the Mott transition in Bose polarons by measuring the intensity of polaron peaks in photoluminescence excitation (PLE) spectra. The PLE signal is a measure of optical absorption. Figure 1c shows that the polaron states vanish with increasing density $n$ and, at high 
$n$, the PLE spectra become step-like due to the step-like 2D density of states. The density-temperature $n-T$ diagram of the lowest-energy polaron state is shown in Fig.~2. The e-h densities $n$ are estimated from the energy shift $\delta E$ of the spatially indirect (interlayer) PL using the 'capacitor' formula $\delta E = 4\pi e^2 d n / \kappa$, where $d$ is the distance between the electron and hole layers, close to the distance between the QW centers~\cite{Yoshioka1990}. The 'capacitor' formula is more accurate at higher $n$~\cite{Choksy2021}. 

Due to the presence of both heavy holes and light holes in GaAs heterostructures, four polaron peaks are observed in the spectra (Fig.~1c). The two lower-energy peaks correspond to ABP and RBP for DXs containing heavy holes (hh); The two higher-energy peaks correspond to ABP and RBP for DXs containing light holes (lh)~\cite{Szwed2024}. Figure 2 shows the $n-T$ diagram for hh ABP state. Similar $n-T$ diagrams for hh RBP, lh ABP, and lh RBP states are shown in SI. 

At low e-h densities $n$, increasing $n$ leads to an enhancement of the polaron density and, in turn, the intensity of the polaron peak (Fig.~2). At high e-h densities, the polaron state vanishes due to the Mott transition (Figs.~1c and 2). Similar behavior is also observed for hh RBP, lh ABP, and lh RBP states, see SI. All the polaron states vanish at high densities due to the Mott transition.

\begin{figure}
\begin{center}
\includegraphics[width=8.5cm]{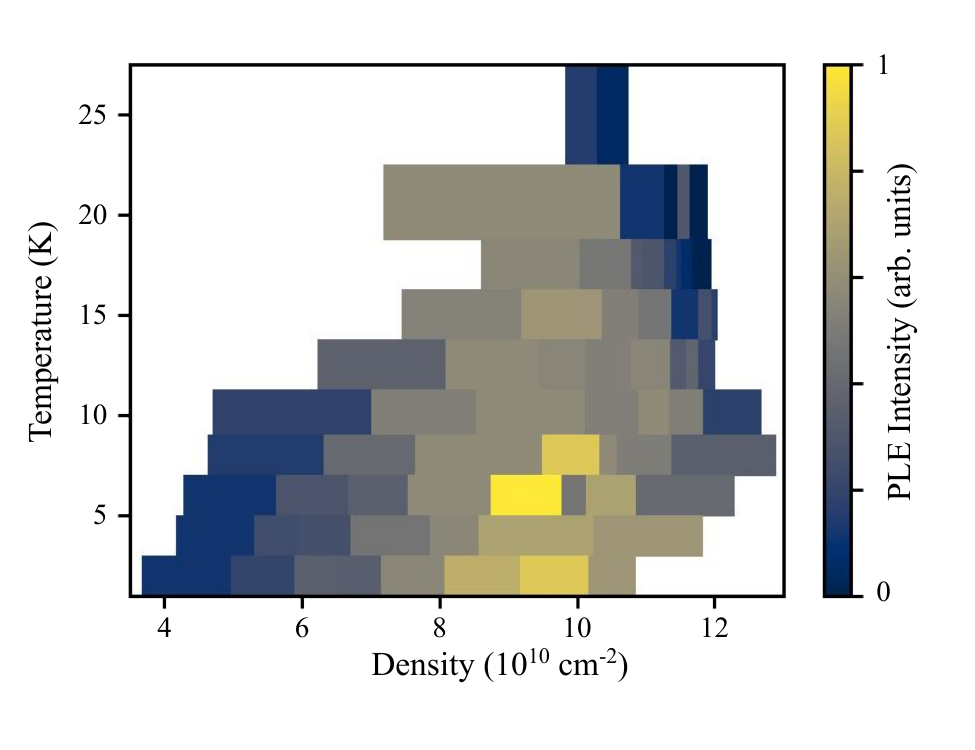}
\caption{$n-T$ diagram for hh ABP polaron state. The intensity of hh ABP peak in PLE spectra is shown vs. density and temperature. At low e-h densities, increasing the density leads to the enhancement of the polaron density and the intensity of the polaron peak. At high e-h densities, the polaron state vanishes due to the Mott transition. 
}
\end{center}
\end{figure}

Figure 3a shows the horizontal cross-sections of the $n-T$ diagram for hh ABP state at different temperatures. The cross-sections of the $n-T$ diagrams for hh RBP, lh ABP, and lh RBP polaron states are similar, see SI. We analyze the drop of the polaron peak intensity at high $n$ and quantify the density of the Mott transition $n_{\rm M}$ by the density at which the polaron peak intensity drops by $e$ times relative to the maximum. These values for $n_{\rm M}$ are shown in Fig.~3b for the four polaron states, hh ABP, hh RBP, lh ABP, and lh RBP, for different temperatures. The Mott transition density $n_{\rm M} \sim 10^{11}$~cm$^{-2}$ for the Bose polaron states (Fig.~3b).

\begin{figure}
\begin{center}
\includegraphics[width=8.5cm]{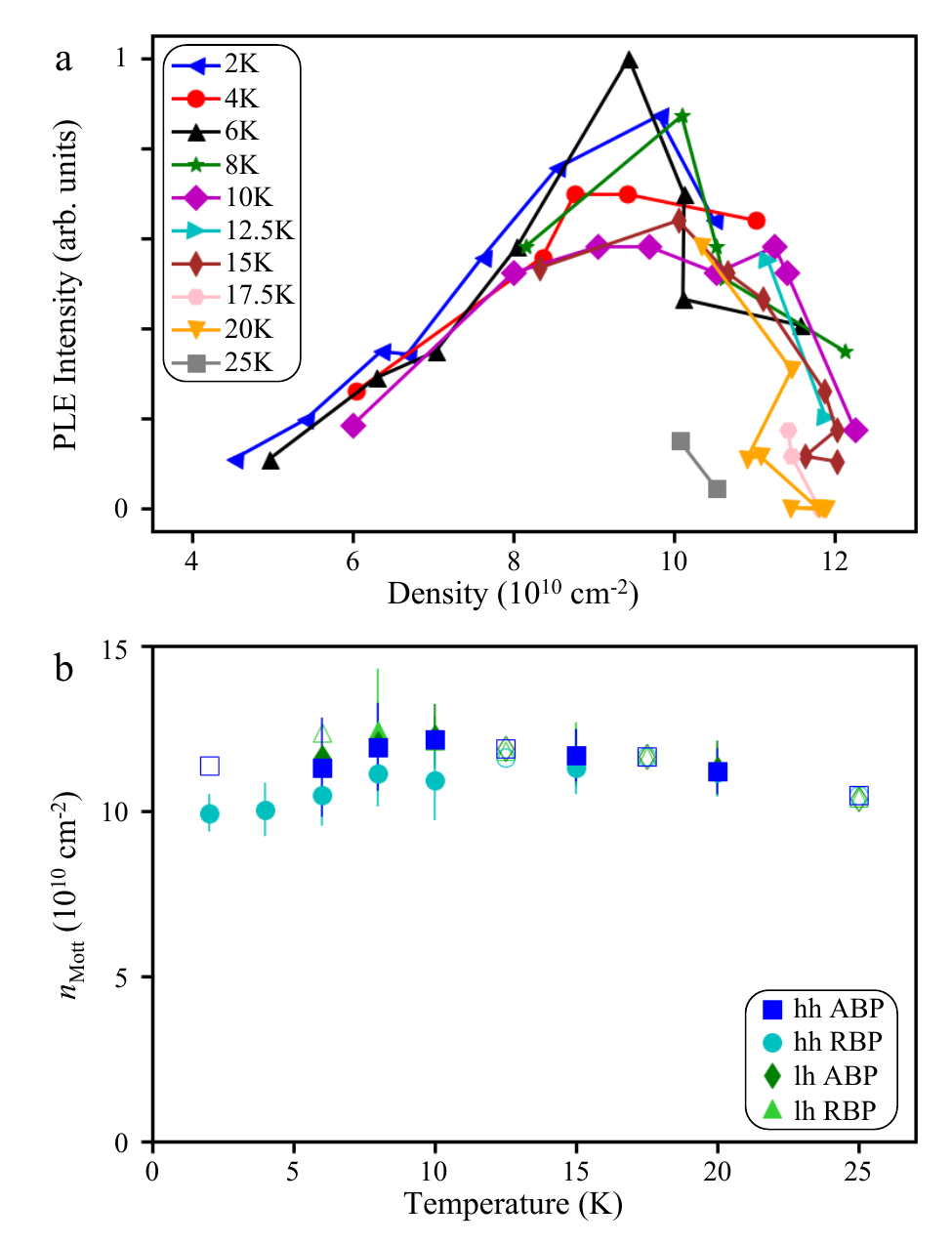}
\caption{The Mott transition in excitonic Bose polarons. (a) hh ABP peak intensity vs. density for different temperatures. At high e-h densities, the polaron state vanishes due to the Mott transition. (b) The density of the Mott transition for the excitonic Bose polarons, $n_{\rm M}$, that is given by the density at which the polaron peak intensity drops by $e$ times relative to the maximum. Open signs correspond to the $n-T$ diagram cross-sections where the reduction of the polaron peak intensity with density was measured whereas the maximum was not measured, the accuracy of these additional data points is lower. The Mott transition density $n_{\rm M} \sim 10^{11}$~cm$^{-2}$ for the Bose polaron states. 
\label{fig:3}}
\end{center}
\end{figure}

The data in Fig.~3b allow determining the Mott transition parameter $n_{\rm M}^{1/2} a_{\rm B}$ for 2D excitonic Bose polarons. This is a 2D analog of the Mott transition parameter for 3D systems $n_{\rm M}^{1/3} a_{\rm B}$~\cite{Mott1961}. To estimate the Mott parameter for 2D excitonic Bose polarons, we calculated the Bohr radius for both DXs and IXs: $a_{\rm B_{DX}} \sim 14$~nm and $a_{\rm B_{IX}} \sim 29$~nm. The calculations are described in SI. For the Mott transition in Bose polarons, the role of the Bohr radius of a mobile composite particle of one kind in a Bose gas of composite particles of another kind and the role of the Bohr radius of the composite particles forming the Bose gas yet need to be clarified. In this work, we give the parameters for both composite particles relevant to the Mott transition: $n^{1/2} a_{\rm B_{DX}}$ and $n^{1/2} a_{\rm B_{IX}}$. Our data for $n_{\rm M}$ (Fig.~3b) and calculated $a_{\rm B_{DX}}$ and $a_{\rm B_{IX}}$ give $n_{\rm M}^{1/2} a_{\rm B_{DX}} \sim 0.4$ and $n_{\rm M}^{1/2} a_{\rm B_{IX}} \sim 0.9$. These values can be compared with the Mott parameter estimated by Mott for 3D systems of pairs of the same kind $n_{\rm M}^{1/3} a_{\rm B} \sim 0.25$~\cite{Mott1961}. The development of theoretical understanding of the Mott transition in Bose polarons is a subject for future work. 

We note also somewhat similar systems of Fermi polarons. A Fermi polaron is a mobile particle of one kind in a Fermi gas of particles of another kind. Fermi polarons in ultracold atomic Fermi gases~\cite{Schirotzek2009, Nascimbene2009, Kohstall2012, Koschorreck2012, Cetina2016} and excitonic Fermi polarons in electron gases in 2D heterostructures~\cite{Sidler2017, Efimkin2017, Chang2018, Fey2020, Glazov2020, Efimkin2021, Tiene2022, Huang2023} are actively studied. Vanishing of polaron states at high densities may be also relevant to Fermi polaron systems.

In summary, we presented evidence for the Mott transition in Bose polarons. For the earlier studied many-body systems, such as arrays of atoms, doped semiconductors, and optically generated electrons and holes, the Mott transition is induced by interaction in a system of pairs of the same kind. The Mott transition for Bose polarons is of a different type. A Bose polaron is a mobile particle of one kind in a Bose gas of particles of another kind. For the Mott transition in polarons, the polaron states vanish with increasing density of the surrounding gas. We presented the observation of this type of the Mott transition and the measurement of the Mott transition parameter $n_{\rm M}^{1/2} a_{\rm B}$ in 2D excitonic Bose polarons. 

We thank Dmitry Efimkin for discussions. The PLE studies were supported by DOE award DE-FG02-07ER46449, the device fabrication by NSF grant DMR 1905478, and the heterostructure growth by Gordon and Betty Moore Foundation grant GBMF9615 and NSF grant DMR 2011750.



\begin{references}
\bibitem{Mott1961}
N.F. Mott, The transition to the metallic state, {\it Philosophical Magazine} {\bf 6}, 287 (1961).

\bibitem{Mott1968}
N.F. Mott, Metal-insulator transition, {\it Rev. Mod. Phys.} {\bf 40}, 677 (1968).

\bibitem{Keldysh1965}
L.V. Keldysh, Y.V. Kopaev, Possible instability of the semimetallic state toward Coulomb interaction, {\it Sov. Phys. Solid State} {\bf 6}, 2219 (1965).

\bibitem{Comte1982}
C. Comte, P. Nozi{\'e}res, Exciton Bose condensation: the ground state of an electron-hole gas -- I. Mean field description of a simplified model, {J. de Phys.} {\bf 43}, 1069 (1982).

\bibitem{Choksy2023}
D.J. Choksy, E.A. Szwed, L.V. Butov, K.W. Baldwin, L.N. Pfeiffer, Fermi edge singularity in neutral electron-hole system, {\it Nat. Phys.} {\bf 19}, 1275 (2023).

\bibitem{Keldysh1968}
L.V. Keldysh, A.N. Kozlov, Collective properties of excitons in semiconductors, {\it Sov. Phys. JETP} {\bf 27}, 521 (1968).



\bibitem{Butov1991}
L.V. Butov, V.D. Kulakovskii, E. Lach, A. Forchel, D. Gr{\"u}tzmacher, Magnetoluminescence study of many-body effects in homogeneous quasi-two-dimensional electron-hole plasma in undoped InGaAs/InP single quantum wells, {\it Phys. Rev. B} {\bf 44}, 10680 (1991).

\bibitem{Kappei2005}
L. Kappei, J. Szczytko, F. Morier-Genoud, B. Deveaud, Direct observation of the Mott transition in an optically excited semiconductor quantum well, {\it Phys. Rev. Lett.} {\bf 94}, 147403 (2005).

\bibitem{Wang2019}
Jue Wang, Jenny Ardelean, Yusong Bai, Alexander Steinhoff, Matthias Florian, Frank Jahnke, Xiaodong Xu, Mackillo Kira, James Hone, X.-Y. Zhu, Optical generation of high carrier densities in 2D semiconductor heterobilayers, {\it Sci. Adv.} {\bf 5}, eaax0145 (2019). 



\bibitem{Hu2016}
Ming-Guang Hu, Michael J. Van de Graaff, Dhruv Kedar, John P. Corson, Eric A. Cornell, Deborah S. Jin, Bose Polarons in the Strongly Interacting Regime, {\it Phys. Rev. Lett.} {\bf 117}, 055301 (2016).

\bibitem{Jorgensen2016}
Nils B. Jørgensen, Lars Wacker, Kristoffer T. Skalmstang, Meera M. Parish, Jesper Levinsen, Rasmus S. Christensen, Georg M. Bruun, Jan J. Arlt, Observation of Attractive and Repulsive Polarons in a Bose-Einstein Condensate, {\it Phys. Rev. Lett.} {\bf 117}, 055302 (2016).

\bibitem{Camargo2018}
F. Camargo, R. Schmidt, J.D. Whalen, R. Ding, G. Woehl, Jr., S. Yoshida, J. Burgd{\"o}rfer, F.B. Dunning, H.R. Sadeghpour, E. Demler, T.C. Killian, Creation of Rydberg Polarons in a Bose Gas, {\it Phys. Rev. Lett.} {\bf 120}, 083401 (2018).

\bibitem{Yan2020}
Zoe Z. Yan, Yiqi Ni, Carsten Robens, Martin W. Zwierlein, Bose polarons near quantum criticality, {\it Science} {\bf 368}, 190 (2020).

\bibitem{Skou2021}
Magnus G. Skou, Thomas G. Skov, Nils B. Jørgensen, Kristian K. Nielsen, Arturo Camacho-Guardian, Thomas Pohl, Georg M. Bruun, Jan J. Arlt, Non-equilibrium quantum dynamics and formation of the Bose polaron, {\it Nat. Phys.} {\bf 17}, 731 (2021).

\bibitem{Grusdt2024}
F. Grusdt, N. Mostaan, E. Demler, L.A.P. Ardila, Impurities and polarons in bosonic quantum gases: a review on recent progress, arXiv:2410.09413 (2024).

\bibitem{Levinsen2019}
J. Levinsen, F.M. Marchetti, J. Keeling, M.M. Parish, Spectroscopic Signatures of Quantum Many-Body Correlations in Polariton Microcavities, {\it Phys. Rev. Lett.} {\bf 123}, 266401 (2019).

\bibitem{Tan2023}
Li Bing Tan, Oriana K. Diessel, Alexander Popert, Richard Schmidt, Atac Imamoglu, Martin Kroner, Bose Polaron Interactions in a Cavity-Coupled Monolayer Semiconductor, {\it Phys. Rev. X} {\bf 13}, 031036 (2023).

\bibitem{Amelio2023}
Ivan Amelio, N.D. Drummond, Eugene Demler, Richard Schmidt, Atac Imamoglu, Polaron spectroscopy of a bilayer excitonic insulator, {\it Phys. Rev. B} {\bf 107}, 155303 (2023).

\bibitem{Szwed2024}
Erik A. Szwed, Brian Vermilyea, Darius J. Choksy, Zhiwen Zhou, Michael M. Fogler, Leonid V. Butov, Dmitry K. Efimkin, Kirk W. Baldwin, Loren Pfeiffer, Excitonic Bose Polarons in Electron-Hole Bilayers, {\it Nano Lett.} {\bf 24}, 13219 (2024).

\bibitem{Cornell2002}
E.A. Cornell, C.E. Wieman, Bose-Einstein condensation in a dilute gas, the first 70 years and some recent experiments, {\it Rev. Mod. Phys.} {\bf 74}, 875 (2002).

\bibitem{Ketterle2002}
W. Ketterle, When atoms behave as waves: Bose-Einstein condensation and the atom laser, {\it Rev. Mod. Phys.} {\bf 74}, 1131 (2002).

\bibitem{Choksy2021}
D.J. Choksy, C. Xu, M.M. Fogler, L.V. Butov, J. Norman, A.C. Gossard, Attractive and repulsive dipolar interaction in bilayers of indirect excitons, {\it Phys. Rev. B} {\bf 103}, 045126 (2021).

\bibitem{Yoshioka1990}
D. Yoshioka, A.H. MacDonald, Double quantum well electron-hole systems in strong magnetic fields, {\it J. Phys. Soc. Jpn.} 59, 4211 (1990).



\bibitem{Schirotzek2009}
Andr{\'e} Schirotzek, Cheng-Hsun Wu, Ariel Sommer, Martin W. Zwierlein, Observation of Fermi Polarons in a Tunable Fermi Liquid of Ultracold Atoms, {\it Phys. Rev. Lett.} {\bf 102}, 230402 (2009).

\bibitem{Nascimbene2009}
S. Nascimb{\'e}ne, N. Navon, K.J. Jiang, L. Tarruell, M. Teichmann, J. McKeever, F. Chevy, C. Salomon, Collective Oscillations of an Imbalanced Fermi Gas: Axial Compression Modes and Polaron Effective Mass, {\it Phys. Rev. Lett.} {\bf 103}, 170402 (2009).

\bibitem{Kohstall2012}
C. Kohstall, M. Zaccanti, M. Jag, A. Trenkwalder, P. Massignan, G.M. Bruun, F. Schreck, R. Grimm, Metastability and coherence of repulsive polarons in a strongly interacting Fermi mixture, {\it Nature} {\bf 485}, 615 (2012).

\bibitem{Koschorreck2012}
Marco Koschorreck, Daniel Pertot, Enrico Vogt, Bernd Fr{\"o}hlich, Michael Feld, Michael K{\"o}hl, Attractive and repulsive Fermi polarons in two dimensions, {\it Nature} {\bf 485}, 619 (2012).

\bibitem{Cetina2016}
Marko Cetina, Michael Jag, Rianne S. Lous, Isabella Fritsche, Jook T.M. Walraven, Rudolf Grimm, Jesper Levinsen, Meera M. Parish, Richard Schmidt, Michael Knap, Eugene Demler, Ultrafast many-body interferometry of impurities coupled to a Fermi sea, {\it Science} {\bf 354}, 96 (2016).



\bibitem{Sidler2017}
Meinrad Sidler, Patrick Back, Ovidiu Cotlet, Ajit Srivastava, Thomas Fink, Martin Kroner, Eugene Demler, Atac Imamoglu, Fermi polaron-polaritons in charge-tunable atomically thin semiconductors, {\it Nat. Phys.} {\bf 13}, 255 (2017).

\bibitem{Efimkin2017}
Dmitry K. Efimkin, Allan H. MacDonald, Many-body theory of trion absorption features in two-dimensional semiconductors, {\it Phys. Rev. B} {\bf 95}, 035417 (2017).

\bibitem{Chang2018}
Yia-Chung Chang, Shiue-Yuan Shiau, Monique Combescot, Crossover from trion-hole complex to exciton-polaron in n-doped two-dimensional semiconductor quantum wells, {\it Phys. Rev. B} {\bf 98}, 235203 (2018).

\bibitem{Fey2020}
Christian Fey, Peter Schmelcher, Atac Imamoglu, Richard Schmidt, Theory of exciton-electron scattering in atomically thin semiconductors, {\it Phys. Rev. B} {\bf 101}, 195417 (2020).

\bibitem{Glazov2020}
M.M. Glazov, Optical properties of charged excitons in two-dimensional semiconductors, {\it J. Chem. Phys.} {\bf 153}, 034703 (2020).

\bibitem{Efimkin2021}
Dmitry K. Efimkin, Emma K. Laird, Jesper Levinsen, Meera M. Parish, Allan H. MacDonald, Electron-exciton interactions in the exciton-polaron problem, {\it Phys. Rev. B} {\bf 103}, 075417 (2021).

\bibitem{Tiene2022}
A. Tiene, J. Levinsen, J. Keeling, M.M. Parish, F.M. Marchetti, Effect of fermion indistinguishability on optical absorption of doped two-dimensional semiconductors, {\it Phys. Rev. B} {\bf 105}, 125404 (2022).

\bibitem{Huang2023}
Di Huang, Kevin Sampson, Yue Ni, Zhida Liu, Danfu Liang, Kenji Watanabe, Takashi Taniguchi, Hebin Li, Eric Martin, Jesper Levinsen, Meera M. Parish, Emanuel Tutuc, Dmitry K. Efimkin, Xiaoqin Li, Quantum Dynamics of Attractive and Repulsive Polarons in a Doped MoSe$_2$ Monolayer, {\it Phys. Rev. X} {\bf 13}, 011029 (2023).

\end{references}
\end{document}


\date{\today}

\title{Supporting Information for\\
Mott transition in excitonic Bose polarons}

\author{E.~A.~Szwed\textsuperscript{a}}
\affiliation{\textsuperscript{a}Department of Physics, University of California San Diego, La Jolla, CA 92093, USA}
\author{B.~Vermilyea\textsuperscript{a}}
\affiliation{\textsuperscript{a}Department of Physics, University of California San Diego, La Jolla, CA 92093, USA}
\author{D.~J.~Choksy\textsuperscript{a}}
\affiliation{\textsuperscript{a}Department of Physics, University of California San Diego, La Jolla, CA 92093, USA}
\author{Zhiwen~Zhou\textsuperscript{a}}
\affiliation{\textsuperscript{a}Department of Physics, University of California San Diego, La Jolla, CA 92093, USA}
\author{M.~M.~Fogler\textsuperscript{a}}
\affiliation{\textsuperscript{a}Department of Physics, University of California San Diego, La Jolla, CA 92093, USA}
\author{L.~V.~Butov\textsuperscript{a}
} 
\affiliation{\textsuperscript{a}Department of Physics, University of California San Diego, La Jolla, CA 92093, USA}
\author{K.~W.~Baldwin\textsuperscript{b}}
\affiliation{\textsuperscript{b}Department of Electrical Engineering, Princeton University, Princeton, NJ 08544, USA}
\author{L.~N.~Pfeiffer\textsuperscript{b}}
\affiliation{\textsuperscript{b}Department of Electrical Engineering, Princeton University, Princeton, NJ 08544, USA}

\begin{abstract}
\noindent
\end{abstract}

\maketitle
\renewcommand*{\thefigure}{S\arabic{figure}}

\subsection{Heterostructure}
\label{sec:heterostructure}

The studied coupled quantum well (CQW) heterostructure is grown by molecular beam epitaxy. The CQW consists of two $15$~$\mathrm{nm}$ GaAs QWs separated by a $4$~$\mathrm{nm}$ Al$_{0.33}$Ga$_{0.67}$As barrier. The CQW is positioned within an undoped $1$-$\mu\mathrm{m}$-thick Al$_{0.33}$Ga$_{0.67}$As layer. An $n^+$ GaAs layer with $n_\mathrm{Si} \sim 10^{18}\,\mathrm{cm}^{-3}$ serves as a bottom gate. The CQW is located $100\,\mathrm{nm}$ above the $n^+$ GaAs layer, closer to the bottom gate, to minimize the effect of fringing electric fields in excitonic devices~\cite{Hammack2006}. $2\,\mathrm{nm}$ of Ti and $7\,\mathrm{nm}$ of Pt evaporated on a $7.5\,\mathrm{nm}$ GaAs cap layer form a top semi-transparent gate. Gate voltage $-2.5\,\mathrm{V}$ creates an electric field normal to the CQW plane driving optically generated electrons and holes to the opposite QWs forming the electron layer and the hole layer (Fig.~1a in the main text). This process is fast, so that the densities of minority particles (electrons in the hole layer and holes in the electron layer) are orders of magnitude smaller than the densities of majority particles (electrons in the electron layer and holes in the hole layer).

\subsection{Optical measurements}

The PLE spectra (Fig.~1c in the main text) probe spatially direct optical absorption within each QW. The PLE spectra are measured $\sim 50\,\mu\mathrm{m}$ away from the laser excitation spot and $\sim 300\,\mathrm{ns}$ after the excitation pulse, where a cold and dense electron hole (e-h) system with temperature close to the lattice temperature is formed~\cite{Choksy2023, Szwed2024}. To enable comparison with prior measurements in e-h bilayers~\cite{Choksy2023, Szwed2024}, we use similar optical excitation and detection protocol, which is described below. The e-h system is generated by a Ti:\,Sapphire laser. An acousto-optic modulator is used for making laser pulses ($800\,\mathrm{ns}$ on, $400\,\mathrm{ns}$ off). A laser excitation spot with a mesa-shaped intensity profile and diameter $\sim 100\,\mu\mathrm{m}$ is created using an axicon. The signal is detected within a $50\,\mathrm{ns}$ window, which is much shorter than the IX lifetime, so that the signal variation during the measurement is negligible~\cite{Choksy2023}. The exciton density in the detection region is close to the density in the excitation spot because the separation is shorter than the IX propagation length and the time delay is shorter than the IX lifetime~\cite{Choksy2023}. The IX PL spectra are measured using a spectrometer with resolution $0.2\,\mathrm{meV}$ and a liquid-nitrogen-cooled CCD coupled to a PicoStar HR TauTec time-gated intensifier. The experiments are performed in a variable-temperature $^4$He cryostat.

\subsection{$n-T$ diagrams for polarons}

Figure S1 shows $n-T$ diagrams for polaron states hh RBP, lh ABP, and lh RBP (similar to $n-T$ diagram for hh ABP in Fig.~2 in the main text). For all polaron states, at low e-h densities, increasing the density leads to the enhancement of the polaron densities and the intensities of the polaron peaks. At high e-h densities, the polaron states vanish due to the Mott transition.

\begin{figure*}
\begin{center}
\includegraphics[width=17.5cm]{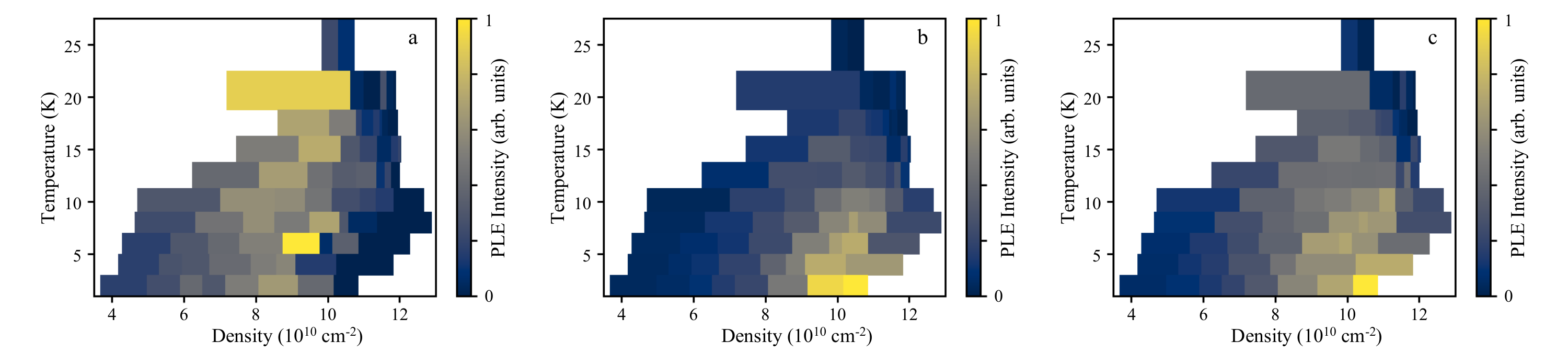}
\caption{$n-T$ diagrams for polaron states hh RBP, lh ABP, and lh RBP (similar to $n-T$ diagram for hh ABP in Fig.~2 in the main text). The intensity of hh RBP (a), lh ABP (b), and lh RBP (c) peak in PLE spectra is shown vs. density and temperature. At low e-h densities, increasing the density leads to the enhancement of the polaron densities and the intensities of the polaron peaks. At high e-h densities, the polaron states vanish due to the Mott transition.}
\end{center}
\end{figure*}

Figure S2 shows the horizontal cross-sections of the $n-T$ diagrams for hh RBP, lh ABP, and lh RBP states at different temperatures (similar to cross-sections of the $n-T$ diagram for hh ABP in Fig.~3a in the main text). 
The polaron states vanish at high e-h densities due to the Mott transition. As described in the main text, we quantify the density of the Mott transition for polarons $n_{\rm M}$ by the density at which the polaron peak intensity
drops by $e$ times relative to the maximum. These values for $n_{\rm M}$ are shown in Fig.~3b in the main text for the polaron states hh ABP, hh RBP, lh ABP, and lh RBP.

\begin{figure*}
\begin{center}
\includegraphics[width=17cm]{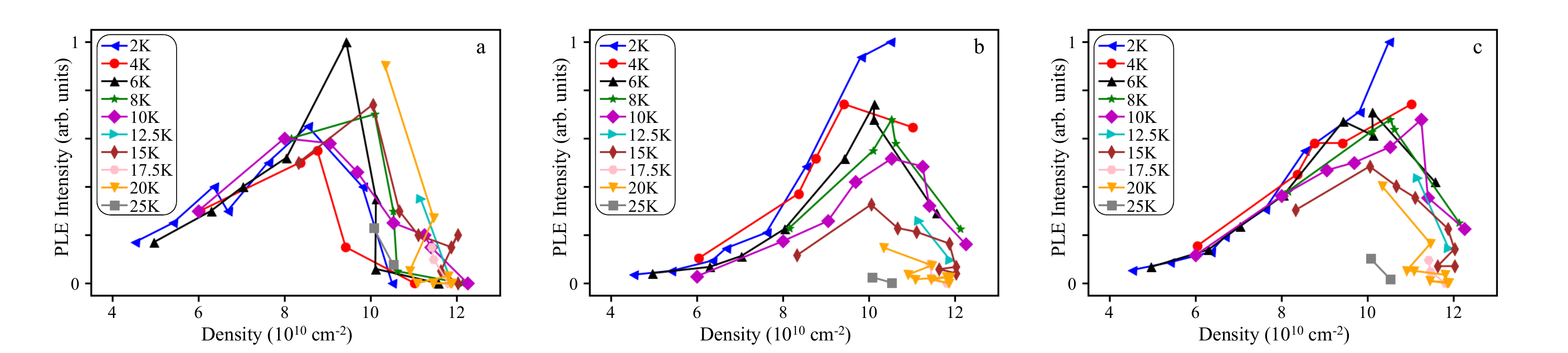}
\caption{The Mott transition in excitonic Bose polarons. (a) hh RBP, (b) lh ABP, and (c) lh RBP peak intensity vs. density for different temperatures. At high e-h densities, the polaron state vanishes due to the Mott transition.
}
\end{center}
\end{figure*}

\begin{figure*}
\begin{center}
\includegraphics[width=17.5cm]{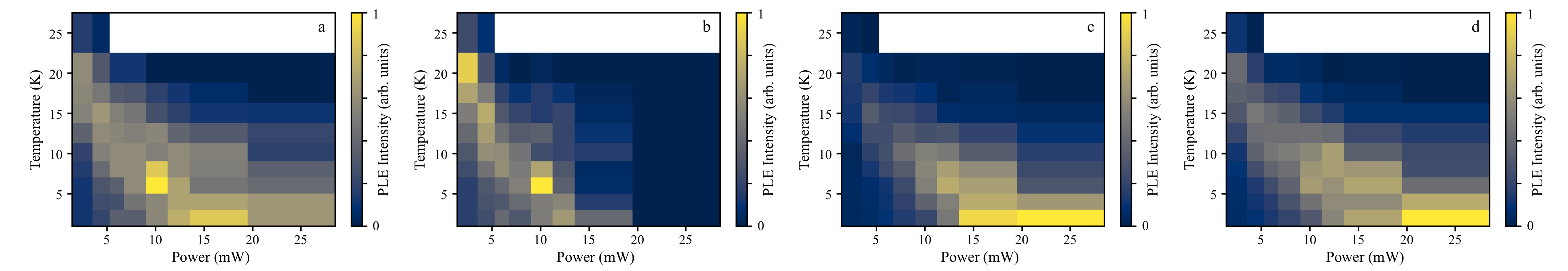}
\caption{$P_{\rm ex}-T$ diagrams for polaron states hh ABP, hh RBP, lh ABP, and lh RBP. The intensity of hh ABP (a), hh RBP (b), lh ABP (c), and lh RBP (d) peak in PLE spectra is shown vs. optical excitation power $P_{\rm ex}$ and temperature. At high $P_{\rm ex}$, the polaron states vanish due to the Mott transition. The $n-T$ diagrams (Fig.~2 in the main text and Fig.~S1) are obtained from the $P_{\rm ex}-T$ diagrams. The e-h densities $n$ for the $n-T$ diagrams are estimated from the shift $\delta E$ of the spatially indirect (interlayer) PL using the 'capacitor' formula as described in the text.}
\end{center}
\end{figure*}

\begin{figure}
\begin{center}
\includegraphics[width=5.5cm]{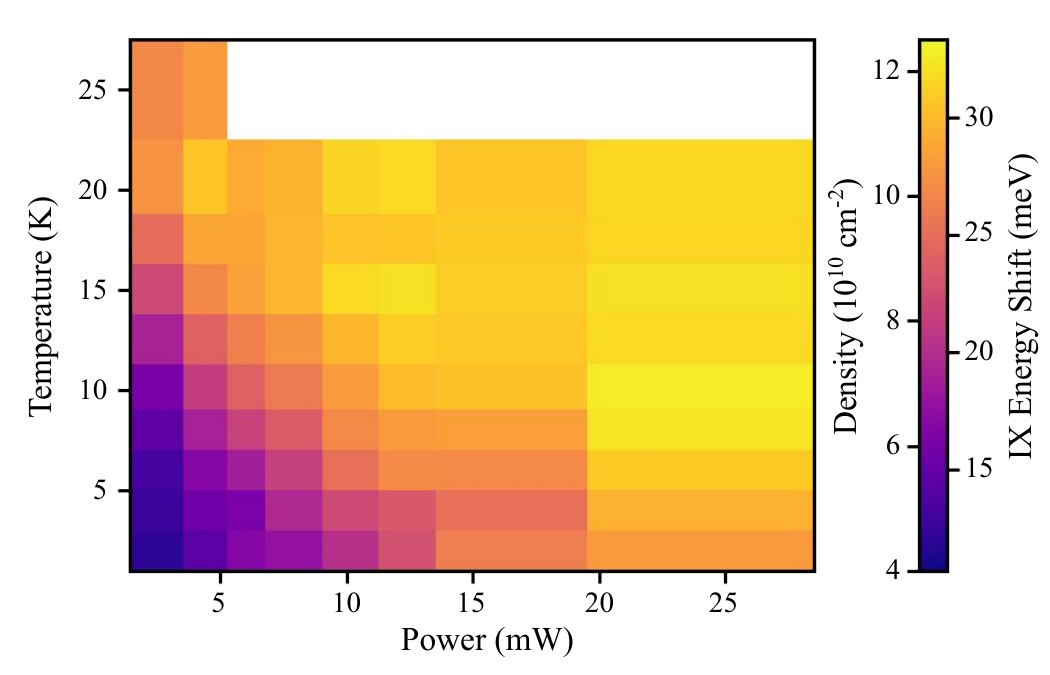}
\caption{The measured shift $\delta E$ of the spatially indirect (interlayer) PL vs. excitation power $P_{\rm ex}$ and temperature. The e-h density $n$ in the e-h bilayer estimated using the 'capacitor' formula, as described in the text, is shown on the color bar. This density is used for the $n-T$ diagrams.}
\end{center}
\end{figure}

\begin{figure*}
\begin{center}
\includegraphics[width=17.5cm]{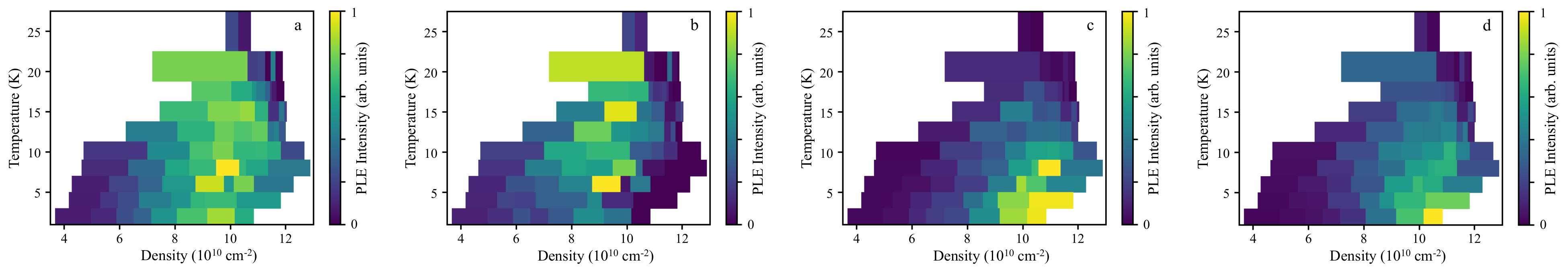}
\caption{$n-T$ diagrams for polaron states hh ABP, hh RBP, lh ABP, and lh RBP similar to $n-T$ diagrams in Fig.~S1 and 2, however for the spectrally integrated intensities of the polaron peaks. The spectrally integrated intensity of hh ABP (a), hh RBP (b), lh ABP (c), and lh RBP (d) peaks in PLE spectra is shown vs. density and temperature. At low e-h densities, increasing the density leads to the enhancement of the polaron densities and the spectrally integrated intensities of the polaron peaks. At high e-h densities, the polaron states vanish due to the Mott transition.}
\end{center}
\end{figure*}

\begin{figure}
\begin{center}
\includegraphics[width=4cm]{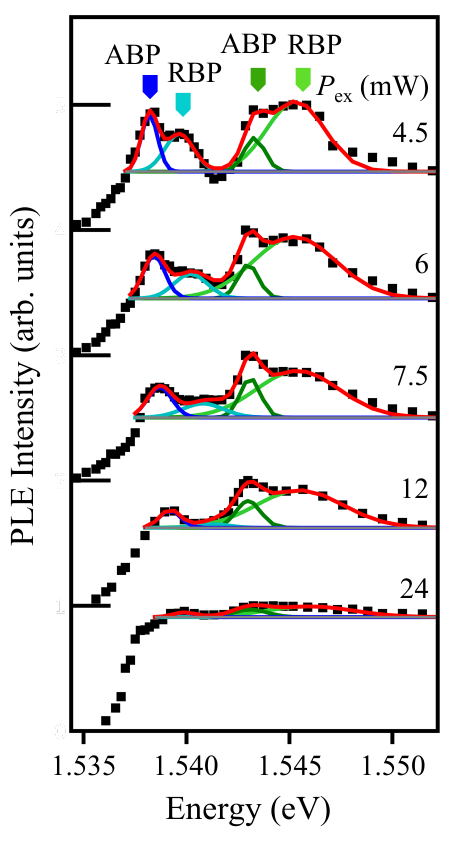}
\caption{Fits of PLE spectra. Gaussian fits for hh ABP is shown in blue, hh RBP in cyan, lh ABP in dark green, and lh RBP in light green. Data points are indicated with black squares, the background intensity with a horizontal gray line, and the sum of spectral fits above the background with a red line. $T=10$K. The peak intensity is normalized.}
\end{center}
\end{figure}

The $n-T$ diagrams (Fig.~2 in the main text and Fig.~S1) are obtained from the $P_{\rm ex}-T$ diagrams (Fig.~S3). The e-h densities $n$ for the $n-T$ diagrams are estimated from the shift $\delta E$ of the spatially indirect (interlayer) PL using the 'capacitor' formula $\delta E = 4\pi e^2 d n / \kappa$, where $d$ is the distance between the electron and hole layers, close to the distance between the QW centers~\cite{Yoshioka1990}. This formula becomes more accurate with increasing $n$~\cite{Choksy2021}. The measured $\delta E$ and estimated e-h density $n$ are shown in Fig.~S4.

The $n-T$ diagrams (Fig.~2 in the main text and Fig.~S1) are qualitatively similar to the $P_{\rm ex}-T$ diagrams (Fig.~S3). Both the $n-T$ and $P_{\rm ex}-T$ diagrams show that the polaron states vanish due to the Mott transition at high $n$ and high $P_{\rm ex}$, respectively. However, the measured $P_{\rm ex}$ and temperature dependence of $\delta E$ and $n$ estimated from these measurements (Fig.~S4) add corrections to the $n-T$ diagrams. In particular, the density $n \sim P_{\rm ex} \tau$ depends on the recombination lifetime $\tau$ and an enhancement of $\tau$ with temperature in the low-density regime of hydrogen-like excitons~\cite{Feldmann1987} contributes to these corrections. The corrections are included to the $n-T$ diagrams where the density is estimated from the measured $\delta E$ as outlined above.

Figure 2 in the main text and Fig. S1 show the intensities of the polaron peaks. The spectrally integrated intensities of the polaron peaks are shown in Fig.~S5. The intensities of the polaron peaks (Figs.~2 and S1) and the spectrally integrated intensities of the polaron peaks (Fig.~S5) display similar density and temperature dependence. Both show that the polaron states vanish at high e-h densities due to the Mott transition.

For the data in Figs.~S1, 2, and S5, the intensities (Figs.~S1 and 2) and the spectrally integrated intensities (Fig.~S5) of the polaron peaks are obtained from the PLE spectra using Gaussian fits as shown in Fig.~S6. At high e-h densities, the polaron peaks vanish and the PLE spectra become step-like due to the step-like 2D density of states. The Gaussian fits for the polaron peaks above the step-like background are shown in Fig.~S6.

\subsection{Exciton binding energies and radii}
\label{sec:dispersion}

\begin{table}[ht]
	\begin{ruledtabular}
		\begin{tabular}{lcc}
			Exciton type & Binding energy (meV) & Radius (nm)
			\\
			\hline 
			IX (e-hh) & $2.99$ & $29.4$
			\\[3pt]
			DX (e-hh) & $8.23$ & $13.6$
			\\[3pt]
			DX (e-lh) & $9.44$ & $11.4$
		\end{tabular}
	\end{ruledtabular}
	\caption{Calculated binding energies and radii of different types of excitons; 'hh' and 'lh' stand for the heavy hole and light hole, respectively.
		\label{tbl:energies}
	}
\end{table}

The calculated binding energies and radii of indirect excitons (IXs)
and direct excitons (DXs) are listed in Table~\ref{tbl:energies}.
This calculation is explained in detail in Ref. \onlinecite{Szwed2024},
so we only briefly outline it here.

We first solved for the single particle states of the QWs \cite{Bastard1986, Vasko1998, Sivalertporn2012}.
This gives the energy-momentum dispersion and wavefunctions of the electron and hole states.
The electron has a parabolic dispersion with effective mass $m_e=0.0665m_0$,
where $m_0$ is the free electron mass.
The heavy hole (hh) and light hole (lh) have non-parabolic dispersions.
To define an effective mass $m_\mathrm{h}$ for the hh,
we fit its dispersion to a parabola
over a range of momenta $0 < k_\bot < a_\mathrm{X}^{-1}$, where
$a_\mathrm{X} = (\kappa \hbar^2/ e^2)(m_\mathrm{e}^{-1} + m_\mathrm{h}^{-1})$ is the exciton Bohr radius.
We found $m_\mathrm{h} = 0.217 m_0 = 3.26 m_\mathrm{e}$.
The lh dispersion is non-monotonic;
for simplicity, we decided to neglect this dispersion altogether,
i.e., to treat the lh mass as infinite.

We used the following expressions for the momentum-space Coulomb interaction potential $\widetilde V(\mathbf k)$ between the electron and hole.
For DXs,
\begin{equation}
\widetilde{V}(\mathbf{k}) =
-\frac{2\pi e^2}{\kappa {k}}\,
\frac{1}{1 + k\rho}\,,
\label{eqn:V_q_approx}\\
\end{equation}
where $\mathbf k$ is the in-plane momentum.
The effective well widths, $\rho = 4.17\,\mathrm{nm}$ for hh and $\rho = 4.38\,\mathrm{nm}$ for lh, 
are determined from the single-particle wavefunctions~\cite{Vasko1998}.
For IXs, we used
$\rho = 0$, i.e., the Coulomb law:
\begin{equation}
\widetilde{V}(\mathbf{k}) = -2\pi \frac{e^2}{\kappa k}\, e^{-kd},
\label{eqn:V_ij_q_Coulomb}\\
\end{equation}
where $d = 19\,\mathrm{nm}$ is the center-to-center layer distance.
Note that intersubband mixing is neglected in our calculations.

We computed the DX and IX binding energies $E_\mathrm{X}$ and ground-state
wavefunctions $\phi_\mathrm{X}(\mathbf{k})$ 
by numerically solving the Wannier equation,
\begin{align}
	[\varepsilon_\mathrm{e}(\mathbf{k}) + \varepsilon_\mathrm{h}(\mathbf{k})]\phi_\mathrm{X}(\mathbf{k})
	+ \Omega^{-1} \sum_{\mathbf{k}'} \widetilde{V}(\mathbf{k} - \mathbf{k}') \phi_\mathrm{X}(\mathbf{k}')
	= -E_\mathrm{X} \phi_\mathrm{X}(\mathbf{k}),
	\label{eqn:H_X}
\end{align}
following Ref.~\onlinecite{Chao1991}.
Here $\mathrm{X} \in \{\mathrm{DX}, \mathrm{IX}\}$ is the exciton type,
$\varepsilon_\mathrm{e,h}(\mathbf{k}) = \hbar^2 \mathbf{k}^2/ 2 m_\mathrm{e,h}$
are the e(h) dispersions,
and $\Omega$ is the area of the system.
The exciton radii $r_\mathrm{X}$ are calculated from the real-space
wavefunctions $\phi_\mathrm{X}(\mathbf r)$:
\begin{equation}
r_\mathrm{X}^2 = \int d^2r\,r^2|\phi_\mathrm{X}(\mathbf r)|^2.
\end{equation}
